\documentclass[12pt]{article}
\usepackage[dvips]{graphicx}

\begin{document}

{\Large{\bf Fresnel tomography: a novel approach to the wave
function reconstruction based on Fresnel representation of
tomograms}}

{\bf S. De Nicola}

{\it Istituto di Cibernetica ``Eduardo Caianiello'' del CNR
Comprensorio ``A.~Olivetti'' Fabbr. 70, Via Campi Flegrei, 34, I-80078
Pozzuoli (NA), Italy}

{\bf R. Fedele}

{\it Dipartimento di Scienze Fisiche, Universit\`{a} ``Federico II''
di Napoli and Istituto Nazionale di Fisica Nucleare, Sezione di
Napoli, Complesso Universitario di Monte Sant Angelo, Via Cintia,
I-80126 Napoli, Italy}

{\bf M. A. Man'ko and V. I. Man'ko}

{\it P. N. Lebedev Physical Institute, Leninskii Prospect, 53,
Moscow 119991 Russia
\\ Email: {\small \bf mmanko@sci.lebedev.ru} }

\begin{abstract}
New type of tomographic probability distribution which contains
complete information on the density matrix (wave function) related
to the Fresnel transform of the complex wave function is
introduced. Relation to symplectic tomographic probability
distribution is elucidated. Multimode generalization of the
Fresnel tomography is presented. Examples of applications of the
present approach are given.
\end{abstract}

{\bf keywords:} {tomographic map, phase space, Wigner function,
Bose--Einstein condensate, Fresnel tomogram.}

\section{Introduction}

Recently symplectic tomography introduced in quantum
optics~\cite{ManciniQSO95} and signal
analysis~\cite{MendesPLA,RitaJRLR} was extended to the problem of
soliton solutions for some nonlinear dynamical equations like
nonlinear Schr\"{o}dinger equation~\cite{tomsol,Gallipoli}
including the study of solitons in Bose--Einstein
condensate~\cite{becsol} described by Gross--Pitaevsky
equation~\cite{G-P}. The name ``symplectic tomography'' is related
to using in the approach standard symplectic group transformations
in phase space (linear canonical transformation in position and
momentum preserving Poisson brackets in classical case and commutator
in quantum case). The symplectic tomographic map associates
with a function (analytic signal, soliton solution, etc.) a
specific probability distribution function (symplectic tomogram)
containing the same information on the signal as the initial
function. It was found in \cite{tomsol,Gallipoli,becsol} that due
to the structure of symplectic tomogram depending on three real
coordinates, one can construct another probability distribution
(Fresnel tomogram). This tomogram depends on two real variables
but information contained in the Fresnel tomogram is sufficient to
reconstruct the initial function for which this tomogram is known.
The inverse formula which permits to make explicit reconstruction
of the initial function (signal, soliton, etc.) was not written
until now and it is one of the aims of this work, namely, to
present the reconstruction formula. Another goal of the work is to
extend the Fresnel tomography to two-particle (multiparticle)
systems to be able to consider in the tomographic representation,
e.g., soliton solutions of nonlinear equations for 2D and 3D
cases.

The paper is organized as follows.

In Section 2, we review the symplectic tomography scheme both for
1D case and multidimensional case. In Section 3, we present the
basic concept of the Fresnel tomography for 1D case while in
Section 4 we present an extension of the Fresnel tomographic
approach to multidimensional problems. In Section 5, we apply our
method to the case of 1D chirped Gaussian wave functions. Finally,
the conclusions are summarized in Section 6.

\section{Symplectic tomograms}

Below we present symplectic tomograms associated with a complex
function $\psi(x)$ describing either a quantum normalized state or
analytic signal or soliton solution of a nonlinear equation. The
tomogram is given by the formula~\cite{MendesPLA}
\begin{equation}\label{1}
w(X,\mu,\nu)=\frac{1}{2\pi|\nu|}\left|\int \psi(y)
\exp\left(\frac{i\mu}{2\nu}\,y^2-\frac{iX}{\nu}\,y\right)\,dy\right|^2.
\end{equation}
One can see that the tomogram which depends on three real variables
has the homogeneity property
\begin{equation}\label{2}
w(\lambda X,\lambda \mu,
\lambda\nu)=\frac{1}{|\lambda|}\,w(X,\mu,\nu).
\end{equation}
The inverse transform which relates the tomogram to the complex
function (density matrix) reads~\cite{RosaPRA}
\begin{equation}
\psi(X)\psi^*(X^\prime)=
\frac {1}{2\pi} \int w(Y,\mu ,X-X^\prime )
\exp\left [i \left (Y-\mu \,\frac {X+X^\prime }{2}\right )\right
]\,d\mu \,dY. \label{3}
\end{equation}

Since one can reconstruct the density matrix in view of the known
symplectic tomogram, all other characteristics like Wigner--Ville
function~\cite{Wigner32,Ville48} can be also expressed in terms of
the symplectic tomogram~\cite{Dariano,FedelePRE}
\begin{equation}\label{4}
W_\psi(q,p)=\int w(X,\mu, \nu)\exp\Big[i(X-\mu q-\nu p)\Big]\,
\frac{dX\,d\mu\,d\nu}{(2\pi)^2}\,,
\end{equation}
where
$$W_\psi(q,p)=\frac{1}{2\pi}\int\psi\left(q+\frac{u}{2}\right)
\psi^*\left(q-\frac{u}{2}\right)e^{-ipu}du.$$
The Wigner function is normalized $W_\psi(q,p)\,dq\,dp=1$.
The symplectic tomogram provides the possibility to find the
optical tomogram simply taking the parameter of the tomogram in
the form $\mu=\cos\theta$, $\nu=\sin\theta$. One arrives at the
formula for the optical tomogram in the form
\begin{equation}\label{6}
w(X,\theta)=\frac{1}{2\pi|\sin\theta|}\left|\int\psi(y)
\exp\left(\frac{i\,\cot\theta}{2} \,y^2-\frac{iX}{\sin\theta}\,y
\right)dy\right|^2.
\end{equation}
Thus optical tomogram can be used in order to reconstruct Wigner
function~\cite{BerBer,VogRis} by means of Radon
transform~\cite{Radon17}.

The generalization of symplectic tomogram to the multidimensional case
is straightforward.

To provide this generalization we introduce the following
notation.

There are three real vectors
$$ \vec X=(X_1,X_2,\ldots, X_N),\quad  \vec\mu=(\mu_1,\mu_2,\ldots, \mu_N),
\quad \vec\nu=(\nu_1,\nu_2,\ldots, \nu_N)$$
and all the components of these vectors vary from $-\infty$ to
$\infty$. The tomogram $w(\vec X,\vec\mu,\vec\nu)$ associated to a
complex function $\psi(\vec x)$, $\vec x=(x_1,x_2,\ldots, x_N)$ reads
\begin{equation}\label{7}
w(\vec
X,\vec\mu,\vec\nu)=\left(\prod_{k=1}^N\frac{1}{2\pi|\nu_k|}\right)
\left|\int\psi(\vec y)\prod_{j=1}^N\exp\left(\frac{i\mu_j}{2\nu_j}\,y_j^2
-\frac{iX_j}{\nu_j}\,y_j\right)\,d\vec y\right|^2.
\end{equation}
The inverse formula has the form
\begin{equation}
\psi(\vec X)\psi^*(\vec X^\prime)=\left(\frac {1}{2\pi}\right)^N
\int w(\vec Y,\vec\mu,\vec X-\vec X^\prime ) \left\{\prod_{k=1}^N
\exp\left[i \left(Y_k-\mu_k \,\frac
{X_k+X_k^\prime}{2}\right)\right] \right\}\,d\vec\mu \,d\vec Y.
\label{8}
\end{equation}

\section{Fresnel tomography in 1D case}

In \cite{tomsol,Fresnel,Piter} the Fresnel tomogram $w_F(X,\nu)$
was introduced
\begin{equation}\label{9}
w_F(X,\nu)=\left|\frac{1}{\sqrt{2\pi i\nu}}\int
\exp\frac{i(X-y)^2}{2\nu} \psi (y)\,dy\right|^2.
\end{equation}
Another version of Fresnel tomography based on using the Fresnel
integral to reconstruct Wigner function of quantum state was suggested
in \cite{SchPRL}.

Since the symplectic tomogram $w(X,\mu,\nu)$ is homogeneous function,
there exists the relation
\begin{equation}\label{10}
w(X,\mu,\nu)=\frac{1}{|\mu|}\,w_F\left(\frac{X}{\mu}\,,\frac{\nu}{\mu}\right).
\end{equation}
This means that if one knows Fresnel tomogram $w_F(X',\nu ')$, the
symplectic tomogram is obtained by the substitution of the arguments
of this function
$$X'\to\frac {X}{\mu}\,,\qquad \nu'\to\frac{\nu}{\mu}$$
and multiplication of the obtained function by the factor
$|\mu|^{-1}$. The formulated relations provide the inverse formula to
reconstruct density matrix by expressing it in terms of Fresnel
tomogram
\begin{equation}
\psi(\vec X)\psi^*(\vec X^\prime)=
\frac {1}{2\pi}\int \frac{1}{|\mu|}\,w_F\left(\frac{Y}{\mu}\,,
\frac{X-X^\prime}{\mu}\right)
\exp\left[i \left(Y-\mu\,\frac {X+X^\prime}{2}\right)
\right]\,d\mu \,dY. \label{11}
\end{equation}
One can also reconstruct the Wigner function expressing it in
terms of Fresnel tomogram
\begin{equation}
W(q,p)=\frac{1}{(2\pi)^2}\int\frac{1}{|\mu|}\,w_F\left(\frac{X}{\mu}\,,
\frac{\nu}{\mu}\right)\exp\,[-i (\mu q +\nu p-X)]\,d\mu \,d\nu \,dX.
\label{12} \end{equation} The inverse formula obtained gives a
possibility to find the Wigner function and the density matrix in the
position representation if one knows the Fresnel tomographic
probability distribution $w_F(X,\nu)\geq 0$, which satisfies the
normalization condition \begin{equation}\label{13} \int
w_F(X,\nu)\,dX=1. \end{equation}

\section{Fresnel tomogram for multipartite system}

Let us introduce the Fresnel tomogram for function $\psi(\vec x)$
depending on several variables. This tomogram can be defined in terms
of symplectic tomogram, i.e.,
\begin{equation}\label{13}
w_F(\vec X,\vec\nu)=w(\vec X,\vec\mu=1,\vec\nu),
\end{equation}
where we have symplectic tomogram and $\vec\mu=\vec 1$ means that the
vector $\vec\mu$ is taken to have all the components equal to one.
Thus the Fresnel tomogram is expressed in terms of the complex
function  $\psi(\vec x)$ as follows:
\begin{equation}\label{14}
w_F(\vec X,\vec\nu)=\left|\prod_{k=1}^N\frac{1}{\sqrt{2\pi
i\nu_k}}\int\psi(\vec y)\left(\prod_{j=1}^N
\exp\frac{i(X_j-y_j)^2}{2\nu_j}\right)d\vec y\right|^2.
\end{equation}
To get the inverse of the above transform, we have to use the
multidimensional analog of relation (\ref{10}) which reads
\begin{equation}\label{15}
w(\vec X,\vec\mu,\vec\nu)=\frac{1}{|\mu_1\mu_2\cdots\mu_N|}\,
w_F\left(\vec X_{\vec\mu}\,,\vec\nu_{\vec\mu}\right),
\end{equation}
where vectors in the argument of Fresnel tomogram have the components
$$\vec X_{\vec\mu}=\left(\frac{X_1}{\mu_1}\,,\frac{X_2}{\mu_2}\,,\ldots
,\frac{X_N}{\mu_N}\right),\qquad
\vec\nu_{\vec\mu}=\left(\frac{\nu_1}{\mu_1}\,,\frac{\nu_2}{\mu_2}\,,\ldots
,\frac{\nu_N}{\mu_N}\right).$$ Using relation (\ref{15}) one can
construct the expression of the density matrix in terms of the Fresnel
tomogram analogous to 1D expression (\ref{11})
\begin{equation}
\psi(\vec X)\psi^*(\vec X^\prime)=\int
\frac{d\vec\mu\,d\vec Y}{(2\pi)^N}\,w_F\left(\vec{Y}_{\vec\mu},
\vec{X}_{\vec\mu}-\vec{X^\prime}_{\vec\mu}\right)\prod_{j=1}^N\frac{1}{|\mu_j|}
\,\exp\left(iY_j-\mu_j\,\frac {X_j+X^\prime_j}{2}\right). \label{16}
\end{equation}

\section{The chirped Gaussian case}

In order to verify the validity of the method, we have numerically
simulated the reconstruction of a 1D complex Gaussian chirped
field (GCF) given by the following normalized form:
\begin{equation}
\psi (x) = \left({2\over \pi \sigma^2}\right)^{1/4}
\exp\left[-{x^2\over\sigma^2} + i\alpha x^2\right]
\label{(1)}
\end{equation}
having  a chirp $\alpha$ and a width determined by the parameter
$\sigma$. The above state is also known as correlated coherent
state~\cite{Kur}. According to Eq. (\ref{3}) the complex field
given by Eq. (\ref{(1)}) can be retrieved from its tomographic
representation $w(X,\mu,x)$ by the following inversion integral:
\begin{equation}
\psi (x)\psi^{*}(0) = {1\over 2\pi}\int\int w(X,\mu,x)
\exp\left[i\left(X-\mu x/2\right)\right]\,dX\,d\mu .\label{(2)}
\end{equation}

The GCF tomogram can be calculated from Eqs. (\ref{1}) and
(\ref{(1)}). The Gaussian distribution can be integrated and the
tomogram can be written in the following form using, according to
the notation of the previous sections, the symbol $\nu$ instead of
$x$
\begin{equation}
w(X,\mu,\nu) = \sqrt{2\sigma^2\over \pi\left[4\nu^2 + \sigma^2
\left(\mu + 2\alpha\nu\right)^2\right]}\,\exp\left\{-{2\sigma^2
X^2\over 4\nu^2 + \sigma^4 \left(\mu +
2\alpha\nu\right)^2}\right\}. \label{(3)}
\end{equation}
Equation (\ref{(3)}) shows that the GCF tomogram is still
characterized by a Gaussian distribution law but its width
$\omega$ is, in general, different from the width of the original
GCF and it can be expressed in terms of the phase-space variables
$\mu$ and $\nu$, chirp parameter $\alpha$, and width $\sigma$,
namely,
\begin{equation}
\omega^2 = {\left(\mu + 2\alpha\nu\right)^2 \sigma^2+4\nu^2\over
2\sigma^2}\,. \label{(4)}
\end{equation}

\begin{figure}
\label{Fig1}
\includegraphics[scale=0.7, viewport=0 0 200pt 300pt]{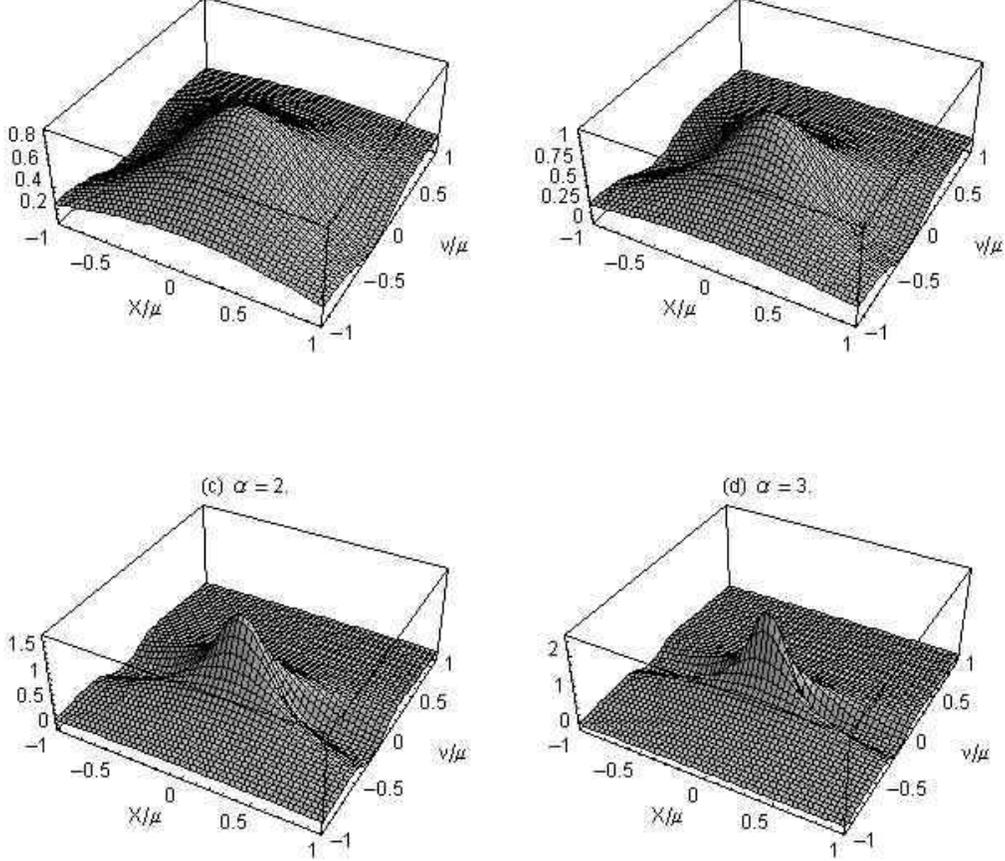}
\caption{3D plot of the
tomogram distribution  $w( X/\mu,1,\nu/\mu)$ of the Gaussian
chirped wave function for a width of the Gaussian $\sigma =1$ and
for increasing values of the chirp parameter $\alpha = 0.5$~(a),
1~(b), 2~(c) and 3~(d).}
\end{figure}

It can be easily verified that the tomogram $w(X,\mu,\nu)$ in Eq.
(\ref{(3)}) coincides with the tomogram of a Gaussian field
without chirp ($\alpha = 0$) computed for the values of the
phase-space variables $\mu^{'} = \mu + 2\alpha\nu$ and
$\nu^{'}=\nu$. We can admit a different transformation of the
phase space, taking into account that the tomographic
representation of the GCF can be also obtained from that of a
Gaussian field without chirp with $\mu^{'}=\mu$  and $\nu^{'}=
\nu\mu/(\mu + 2\alpha\nu)$. In both instances, the effect of the
chirp is that of shifting one of the two values of the phase-space
variables while leaving unaltered the other one. To illustrate the
behaviour of the GCF tomogram, we have displayed in Fig. 1 the 3D
distribution of the tomographic map $w(X/\mu,1,\nu/\mu)$ for
increasing values of the chirp parameter. This representation of
the tomogram in terms of two parameters $X/\mu$ and $\nu/\mu$
rather than three is quite convenient since it allows one a clear
visualization  of the tomographic map in the 2D phase-space plane
($X/\mu$, $\nu/\mu$). Indeed, recalling the general homogeneity
property of the tomogram we can obtain complete dependence of the
tomogram from its three space variables by virtue of Eq.
(\ref{10}):
\begin{equation}
w(X,\mu,\nu) = {1\over |\mu|}\,w(X/\mu,1,\nu/\mu). \label{(5)}
\end{equation}

\begin{figure}
\label{Fig2}
\includegraphics[scale=0.7, viewport=0 0 200pt 300pt]{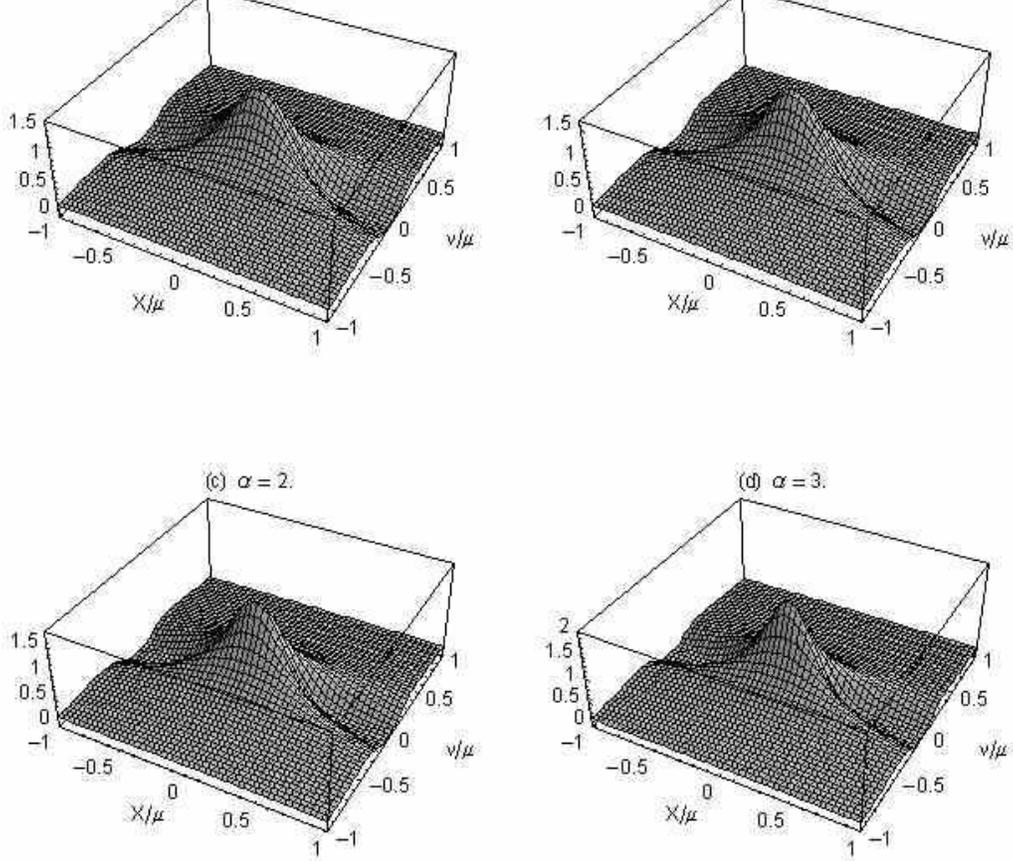}
\caption{3D plot of the tomogram distribution  $w(
X/\mu,1,\nu/\mu)$ of the Gaussian chirped wave function for a
width of the Gaussian $\sigma =0.5$ and for increasing values of
the chirp parameter $\alpha = 0.5$~(a), 1~(b), 2~(c) and 3~(d).}
\end{figure}

According to the Fresnel-based interpretation of the optical
tomogram discussed in the previous section, $w(X/\mu,1,\nu/\mu)$
represents the GCF intensity at distance $\nu$ and Eq. (\ref{(5)})
tells us that the general dependence of the tomogram of its three
real variables $X$, $\mu$, and $\nu$ can be recovered from a set
of measurements of the intensity distributions of the propagated
GCF performed at different distances with a varying scale factor
$1/|\mu|$.  Equation (\ref{(5)}) can be verified immediately from
the defining expression of the GCF tomogram given by
Eq.~(\ref{(3)}). Figure 1 shows clearly that tomographic
distributions shrink in the phase space with increasing the chirp
parameter of the Gaussian field. The numerical results have been
obtained  for a GCF of width $\sigma =1$. In Fig. 2 we display the
plots of the tomograms for the same set of values of Fig. 1 except
that now the width of the GCF is $\sigma = 0.5$. As can be seen by
comparing Figs. 1 and 2, the shrinking of the 3D tomographic
distribution in the phase space with increasing the chirp
parameter is less pronounced when reducing the GCF width. A
computationally efficient method for retrieving the wave function
makes direct use of the inversion integral given by Eq.
(\ref{(2)}) which allows one to determine $\psi (x)$ up to the
complex constant quantity $\psi (0)^{*}$. The method employs the
fast Fourier transform (FFT) algorithm for computing the
two-dimensional Fourier transform
$\hat{w}(\omega_X,\omega_\mu,\nu)$ of the tomogram, namely,
\begin{equation}
\hat{w}(\omega_X,\omega_\mu,\nu) = {1\over
2\pi}\int_{-\infty}^{\infty}
\int_{-\infty}^{\infty}w(X,\mu,\nu)\exp\left[i\left(\omega_X X+
\omega_\mu \mu\right)\right]\,dX d\mu, \label{(23)}
\end{equation}
where $\omega_X$ and $\omega_\mu$ are the spatial frequencies
corresponding to the phase-space variables $X$ and $\mu$ of the
tomogram. The inversion integral is a particular case of the
two-dimensional Fourier transform of the tomogram given by Eq.
(\ref{(23)}). In fact, we have
\begin{equation}
\psi(\nu)\psi(0)^{*} = \hat{w}(1,-\nu/2,\nu). \label{(24)}
\end{equation}
Equation (\ref{(24)}) shows that we can determine the complex
quantity $\psi(\nu)\psi(0)^{*}$ from the values of two-dimensional
Fourier transform of the tomogram at frequencies $\omega_X$ and
$\omega_\mu$, where $\nu$ is allowed to vary in the range along
which the tomographic measurements are performed. In principle,
this procedure works well for reconstructing a wave field, if we
have a sufficient number of sampled values of the tomographic
distribution $w(X,\mu,\nu)$, i.e., if the sampling rate is at
least larger than twice the Nyquist rate. In this case, the
discrete two-dimensional Fourier transform
$\hat{w}(r\Delta\omega_X,s\Delta\omega_\mu)$ corresponding to the
continuous Fourier integral given by Eq. (\ref{(23)}) can be
written in the following form:
\begin{equation}
\hat{w}(r\Delta\omega_X,s\Delta\omega_\mu) = \Delta
X\Delta\mu~\sum_{n=0}^{N-1} \sum_{m=0}^{N-1}~w(n\Delta X,m\Delta
\mu,\nu)\exp\left[-2\pi i\left( {rn\over N}+{sm\over
M}\right)\right], \label{(25)}
\end{equation}
where $r,s$ and $n,m$ are integer numbers,  $\Delta X$ and
$\Delta\mu$ are the sampling intervals in the ($X,\mu$) phase
space and $\Delta\omega_X$ and $\Delta\omega_\mu$ are the
corresponding sampling intervals in the frequency space.
Equation~(\ref{(25)}) allows one to compute a matrix of $N^2$
complex numbers $\hat{w}(r\Delta\omega_X,s\Delta\omega_\mu)$ from
the sampled values $w(n\Delta X,m\Delta \mu,\nu)$ of the tomogram
of the wave function via the discrete two-dimensional fast Fourier
transform (FFT) algorithm. It should be remarked that in order to
obtain accurate reconstruction of the wave function $\psi (\nu)$,
the two-dimensional FFT algorithm needs to be repeatedly applied
to every two-dimensional tomographic distribution corresponding to
each $\nu$ value in the considered range and, if high accuracy is
required, even this FFT-based reconstruction method  becomes
computationally expensive. However, in many cases of interest,
symmetry consideration allows one to simplify somewhat the task
through the consideration of a reduced set of sampled values of
the tomogram. In the GCF case, the two-dimensional Fourier
transform $\hat{w}(\omega_X,\omega_\mu,\nu)$ can be readily
obtained in the following form:
\begin{equation}
\hat{w}(\omega_X,\omega_\mu,\nu) = \sqrt{2\over
\pi\sigma^2\omega_X^2} \exp\left[-{\nu^2\over
2}\left({\omega_X\over\sigma}\right)^2-{2\over \sigma^2}
\left({\omega_\mu\over\omega_X}\right)^2-2i\alpha\omega_\mu
\nu\right]. \label{(26)}
\end{equation}
By using Eq. (\ref{(26)}) it can be easily verified that
Eq.~(\ref{(24)}) gives correctly $\psi (\nu)\psi (0)^{*}$ with the
wave function $\psi (\nu)$ given by the Gaussian chirped field in
Eq. (\ref{(1)}).

\section{Conclusions}

We summarize the main results of the paper.

We obtained the inverse formula which in explicit form
reconstructs the wave function, density matrix and Wigner function
from the known Fresnel tomographic probability distribution
function. The Fresnel tomogram can be used to measure the phase of
radiation by measuring the intensity of the radiation.

We introduced Fresnel tomogram for multimode system and found
reconstruction formula for wave function and density matrix in
terms of the tomograms for this case too. An example of soliton
solution used also in analyzing some states of Bose--Einstein
condensate can be studied from the viewpoint of Fresnel
tomography~\cite{Gallipoli,becsol}.

The symplectic tomography~\cite{ManciniQSO95} has as a partial
case the optical tomography procedure. In this work, we clarified
that there exists another important partial case of the symplectic
tomography which is named Fresnel tomography. This mean that there
exists a connection of Fresnel tomography with well-developed
optical tomography scheme. The tomographic probability of optical
tomography scheme is connected with Fresnel tomogram by the
relation:
$$w(x,\cos\theta,\sin\theta)=\frac{1}{|\cos\theta|}\,w_F\left(
\frac{X}{\sin\theta}\,,\cot\theta\right).$$ One can generalize
this formula to multidimensional case.

\section*{Acknowledgments}

This study was supported by Universit\'{a} ``Federico II'' di
Napoli and the Russian Foundation for Basic Research under Project
No.~03-02-16408. M.A.M. thanks Organizers of the III International
Workshop ``Nonlinear Physics: Theory and Experiment'' for kind
hospitality and the Russian Foundation for Basic Research for
Travel Grant No.~04-02-26662.

\end{document}